\UseRawInputEncoding
\documentclass[trackchanges, twocolumn]{aastex701}
\usepackage{xcolor}
\usepackage{amsmath}
\usepackage{xspace}

\usepackage{url}
\usepackage{etoolbox}
\makeatletter
\patchcmd{\@makefntext}{\parindent 1em}{\def\@footnoterule{\kern-3\p@ \hrule \@width \columnwidth \kern 2.6\p@}\parindent 1em}{}{}
\makeatother

\newcommand{\Mfit}{$M_{fit}$\xspace}
\newcommand{\Mtrue}{$M_{true}$\xspace}
\newcommand{\Efit}{$E_{fit}$\xspace}
\newcommand{\Etrue}{$E_{true}$\xspace}
\newcommand{\Xfit}{$X_{fit}$\xspace}
\newcommand{\Xtrue}{$X_{true}$\xspace}
\newcommand{\dbar}{$\bar{d}$\xspace} 

\begin{document}

\title{Quantifying Element Importance for Mass Recovery from Population III Supernova Yield Fits}

\author[0009-0007-5954-7915]{Zhongyuan Zhang}
\email{zzhang253@uchicago.edu}
\affiliation{Department of Astronomy \& Astrophysics, University of Chicago, 5640 S Ellis Avenue, Chicago, IL 60637, USA}

\author[0000-0002-4863-8842]{Alexander~P.~Ji}
\affiliation{Department of Astronomy \& Astrophysics, University of Chicago, 5640 S Ellis Avenue, Chicago, IL 60637, USA}
\affiliation{Kavli Institute for Cosmological Physics, University of Chicago, Chicago, IL 60637, USA}
\affiliation{NSF-Simons AI Institute for the Sky (SkAI), 172 E. Chestnut St., Chicago, IL 60611, USA}
\email{alexji@uchicago.edu}

\author[0000-0003-4479-1265]{Vinicius M.\ Placco}
\affiliation{NSF NOIRLab, Tucson, AZ 85719, USA}
\email{vinicius.placco@noirlab.edu}

\author[0000-0002-3211-303X]{Sanjana Curtis}
\affiliation{Department of Physics, Oregon State University, 301 Weniger Hall, Corvallis, OR 97331-6507}
\email{sanjana.curtis@oregonstate.edu}

\begin{abstract}
Massive Population III stars are currently not observed, but their initial mass function (IMF) can be inferred through stellar archaeology: fitting core-collapse supernova yield models to elemental abundances of low-mass, long-lived metal-poor stars. While prior work demonstrates that yield fitting can recover progenitor properties, it remains unclear which measured elements most control mass recovery quality and what level of IMF precision is achievable for a measured element set. We perform a systematic study of element importance for progenitor mass recovery. Using the Heger $\&$ Woosley (2010) yield grid, we generate mock observations, fit the initial mass, and evaluate the typical performance on the fractional mass recovery. Add/remove-one-element experiments and comparisons among different baseline element sets are used to rank elements by importance. We find that the most important elements for accurate mass recovery are C, N, Na, and K, with O, Al, Co, and Ni consistently improving performance when available. Overall, with currently measurable elements from high-resolution spectroscopy, stellar archaeology can deliver practical Population III IMF constraints assuming the core-collapse supernova yield models provide a good representation of stellar evolution in the early universe.

\end{abstract}


\section{Introduction} 

Population III (Pop III) stars are the first stars that signal the end of cosmic dark ages \citep{BARKANA_2001}. They have zero metallicity and must have existed, but have not yet been directly observed \citep{Bromm_2013}. The stellar initial mass function (IMF) describes the statistical distribution of stellar masses at birth \citep{Klessen_2023}. In metal-free conditions, the most robust theoretical prediction of Pop III stars is that the IMF is top-heavy \citep{Bromm_1999, Abel_2002, Bromm_2002, Bromm_Volker_2002, OShea_2007, Yoshida_2008, McKee_2008, Greif_2010, Hosokawa_2011, Hosokawa_2012, Stacy_2012, Kippenhahn_2012, Hirano_2014, Susa_2014, Frebel_Norris_2015, Klessen_2023, Sharda_202507}. 

The stellar mass of Pop III stars governs the ionizing spectra and feedback \citep{Schaerer_2002, Tumlinson_2004, Karlsson_2013, Bromm_2013}, remnant outcomes, lifetimes, and nucleosynthetic yields \citep{Frebel_Norris_2015, Klessen_2023}, so constraining the Pop III IMF is central to studying cosmic dawn. 

Pop III stars are also the first source of hydrogen-ionizing photons and initialize cosmic reionization \citep{Tumlinson_2004, Karlsson_2013, Bromm_2013}. Because the ionizing efficiency rises steeply with stellar mass and peaks below the classic very massive star (VMS; $M > 140 M_\odot$) regime, trimming the IMF’s low-mass end maximizes photons per baryon. Therefore, better constraints on the IMF, especially its lower-mass cutoff, translate directly into tighter predictions for when and how strongly reionization proceeds \citep{Tumlinson_2004}. 

The IMF also regulates end-state demographics: Pop III compact binaries can be sources of gravitational waves \citep{Belczynski_2004, Kulczycki_2006, Kowalska_2012, Kinugawa_2014, Kinugawa_2016}, and different assumptions about the Pop III IMF lead to materially different predictions for the formation efficiency, masses, and merger rates of compact binaries; tighter empirical constraints on the IMF would therefore constrain predicted GW detection rates and mass distributions \citep{Kinugawa_2014}. 

Population III stars are also the first sources of heavy-element enrichment \citep{Heger_2002, Heger_2010, Tumlinson_2004, Bromm_2013, Nomoto_2013, Frebel_Norris_2015}. Because stellar mass determines stellar fates, the IMF sets the integrated nucleosynthetic yields of the first stellar population and thus the onset of the Universe’s chemical enrichment, making it central to stellar archaeology \citep{Frebel_Norris_2015}. 

In summary, the Pop III IMF shapes the chemical record in metal-poor stars and is therefore a key input to interpreting observations and modeling the roles of the first stars in the formation, evolution, and chemical enrichment of early stars and galaxies \citep{Frebel_Norris_2015, Klessen_2023}.

Because Population III stars remain undetected directly, their properties must be inferred indirectly \citep{Cayrel_2004, Beer_Christlieb_2005, Frebel_Norris_2015, bonifacio2025metalpoorstars}. The most metal-poor stars, especially extremely metal-poor (EMP; [Fe/H] $< -3$) and ultra metal-poor (UMP; [Fe/H] $< -4$) stars, may be second-generation stars whose surface abundances contain critical information about Population III supernova explosions \citep{Beer_Christlieb_2005, Frebel_Norris_2015}. Because of their low metallicity, an assumption is often made that the heavy elements in each EMP or UMP star originate from only one Pop III star. Therefore, fitting detailed chemical abundances of EMP or UMP stars with Pop III yield models is a way to recover the initial progenitor mass of the Pop III star \citep{Hartwig_2023, Hartwig_2024}. Fitting multiple EMP or UMP stars may give a constraint on the IMF.

Pop III yield fits have been applied to both individual stars \citep{Ishigaki_2014, Frebel_2015, Bessell_2015, Nordlander_2017, Nordlander_2019, Magg_2020, Skuladottir_2021, Placco_2021, Placco_2024, Placco_2025} and multiple stars \citep{Hansen_2011, Tominaga_2014, Placco_2015, Placco_2016, Fraser_2017, Ishigaki_2018, Magg_2020, Jiang_2024, Jiang2025}. Among multiple star studies, \cite{Fraser_2017, Ishigaki_2018} and \cite{Jiang_2024} used the mass recovered from Pop III yield fits to derive the Pop III IMF. Because the IMF inferences hinge on the reliability of the recovered progenitor masses, the next step is to test how stable these fits are and which elements carry the decisive information.

Prior works have been done to assess the robustness of this practice and which elements carry most important information for recovering progenitor mass, such as perturbing individual measurements to gauge sensitivity \citep{Frebel_2015, Placco_2015}, removing one or more elements to test redundancy and stability \citep{Frebel_2015, Placco_2015}, comparing the $\chi^2$ gap between the best and second-best models as a robustness metric \citep{Placco_2015}, and introducing a quantitative metric that reflects how uncertainty of each element abundance affects progenitor derivation \citep{Jiang2025}. Some of these Pop III fitting studies have highlighted the importance of specific elements: \cite{Hansen_2011} highlights Na, Mg, Al, Sc, and Mn as informative diagnostics of distinct model features. \cite{Ishigaki_2014} shows that both SN and HN channels can reproduce measured C, Mg, and Ca, and that additional constraints on mass and energy benefit from O and iron-peak elements such as V, Mn, Co, and Cu. \cite{Tominaga_2014} demonstrates that ratios such as [C/O], [(Na, Al)/Mg], and [Mg/Fe] help constrain mass, especially the ratio between Na/Mg/Al, and underscore the value of measuring as many abundances as possible. \cite{Bessell_2015} finds that C and O vary strongly at fixed explosion energy (implying mass sensitivity) and that best-fit mass can hinge on upper limits in Na and Al. \cite{Placco_2015} reports a strong dependence of fits on C and N, with the presence or absence of N having strong impacts, and no clear relation between progenitor mass and Fe. \cite{Ishigaki_2018} identifies [(C+N)/O] as particularly diagnostic; ratios among Na, Mg, and Al remain useful, and for SN (but not HN) models, Na, Mg, Si, Co, and Zn decrease monotonically with progenitor mass. Finally, \cite{Jiang2025} finds that C, Mg, Mn, Fe, and Co exert the strongest influences in their sensitivity analysis, implicating iron-peak elements as especially informative for constraining EMP progenitors.

Building on these studies, our motivation is to provide a systematic assessment of element-wise information content in Pop III yield fitting. Prior work has typically relied on all elements available in a given spectrum, implicitly assuming that arbitrary combinations return trustworthy progenitor masses. We instead pose two questions: for ensemble IMF inference across multiple stars, which elements are necessary to observe to secure robust mass constraints; and, given realistic measurement uncertainties and incomplete element sets, how well can the progenitor mass be recovered at all? By quantifying the individual and joint contributions of different elements and identifying minimal informative subsets, we aim to establish principled targets for abundance measurements and to determine the achievable precision of mass recovery.

Accordingly, we address the question in a controlled setting: beginning with yield models rather than observations, we generate mock abundance vectors, perturb them within realistic uncertainties, and refit to the same grid to quantify mass-recovery fidelity. Because the progenitor mass is known for each mock observation, we can directly compare recovered best-fit masses to the truth. We then define a metric that compresses each fit into a numerical score and aggregates performance across the full mass-energy grid. By repeating the calculations while adding or removing individual elements in smaller subsets, we determine each element’s importance. Finally, we map the metric to expected errors in IMF inference from stellar samples and, given target precision thresholds, provide recommendations on which elements are necessary or most valuable to measure. This work is outlined as follows: Section \ref{Methods} describes the yield grid and our end-to-end mock-and-refit framework, including the element sets considered, the definition of the mass recovery quality metric, and the summary statistic used to compare performance across element combinations. Section \ref{Results} presents the main results, including element importance rankings and a comparative evaluation of representative and additional element sets, with the complete set of tests provided in the Appendix. Section \ref{Conclusion} summarizes the main conclusions and outlines several directions for extending this framework.

\section{Methods} \label{Methods}
\subsection{Mock Observation Generation}
In this study, we adopt the \citet{Heger_2010} theoretical Pop III yields (HW10) models parameterized by progenitor mass ($10-100 M_\odot$), kinetic energy ($0.3-10$ B; 1B = 1Bethe = $10^{51}$ erg), and mixing ($0 -0.25$ dex). The grid contains 16,800 models.
Each model provides elemental yields from hydrogen to zinc. We restrict attention from carbon through zinc ($Z=6-30$), as abundances of heavier elements are negligible for our purposes. 

\begin{figure}[h]
    \centering
    \includegraphics[width=\linewidth]{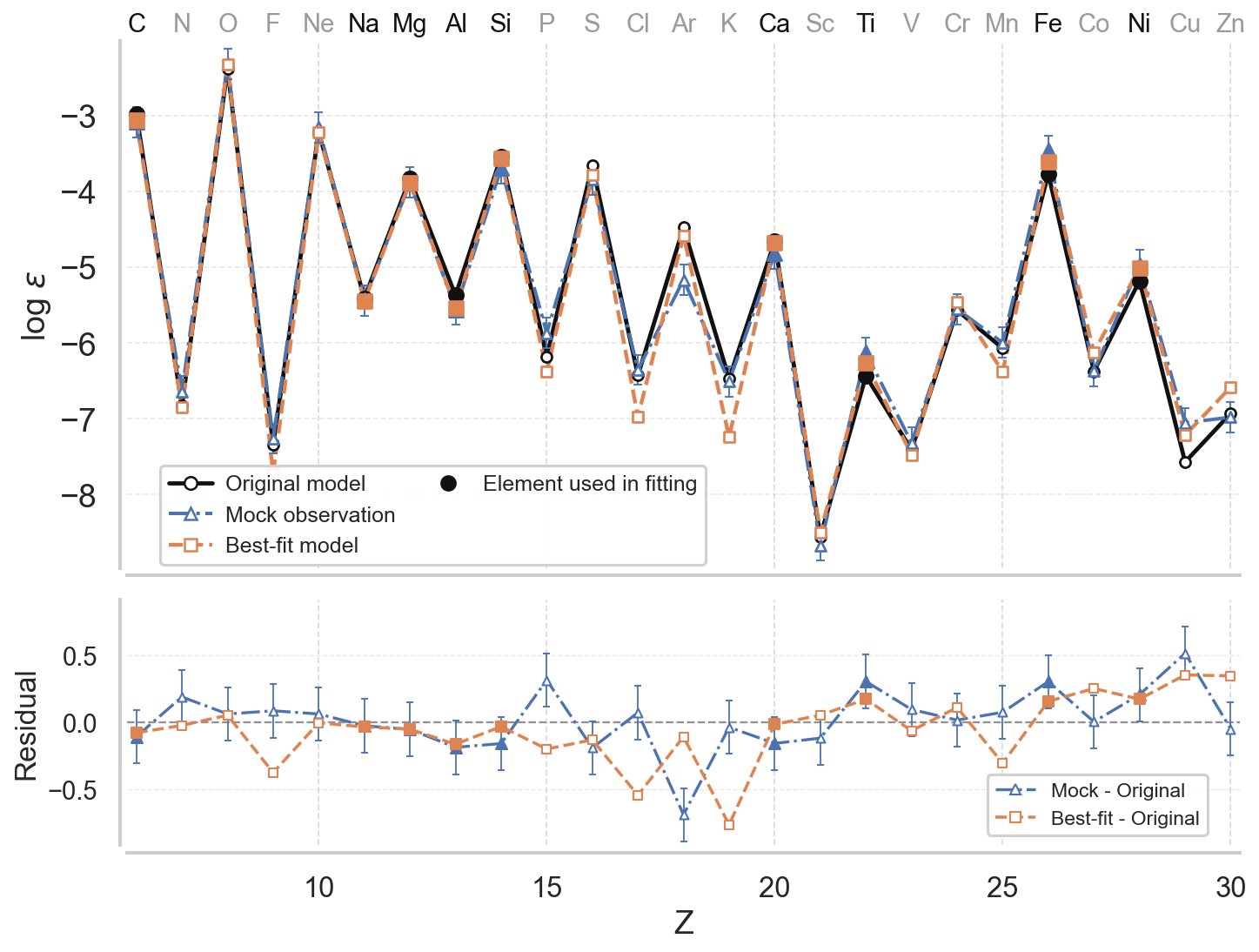}
    \caption{Example of mock observation generation and mass recovery for a HW10 model. Top panel: black circles show the original model abundances in $\log \varepsilon$ with $M_{\rm true}=17.4M_{\odot}$, $E_{\rm true}=2.4$B, and $X_{\rm true}=0.0$; blue triangles show one mock observation generated by adding independent Gaussian noise with $\sigma=0.2$ dex to each element (error bars indicate $1\sigma$); orange squares show the best-fit HW10 model returned by a full-grid $\chi^2$ search with $M_{\rm fit}=18.2M_{\odot}$, $E_{\rm fit}=2.4$B, and $X_{\rm fit}=0.251$. In this example, the fit uses the medium coverage baseline defined in Section \ref{sec 2.2}. Elements included in the fit are indicated by black element labels and filled markers, while excluded elements are shown with gray labels and open markers. Bottom panel: residuals relative to the original model, shown for the mock observation (blue; Mock$-$Original) and the best-fit model (orange; Best-fit$-$Original). Repeating this procedure over 100 independent noise realizations per model yields the distributions of recovered parameters used in the metric analysis in Section \ref{Sec 2.3}.}
    \label{fig:0}
\end{figure}

Mock observations are generated from HW10 models. For each specific star model, we label its mass, energy, and mixing as \Mtrue, \Etrue, and \Xtrue. Element abundances are converted to $\log \varepsilon$, and independent Gaussian noise is added with mean equal to the model abundance and standard deviation $\sigma$ (the “noise level”), which is assumed to be the same for all elements. 
Each mock observation, together with a specified element list, is passed to a fitting code that searches the full HW10 grid and returns the model minimizing the $\chi^2$ mismatch between the model and the mock observation for the chosen elements. 
Figure \ref{fig:0} illustrates the procedure for a HW10 model: starting from the original model abundance pattern (black), we generate a mock observation by perturbing each element abundance with independent Gaussian noise of standard deviation $\sigma = 0.2$ (blue). We then fit this mock observation and find the best-fit HW10 model (orange).
We denote the returned best-fit model parameters by \Mfit, \Efit, and \Xfit. For each original model, we generate 100 independent mock realizations and thus obtain 100 values of \Mfit, \Efit, and \Xfit.

The degrees of freedom of the whole process are the noise level $\sigma$ and the list of elements used in fitting. In this paper, we focus on mass and leave energy to future work. From exploratory tests, we found that mixing is not retrievable, even with a relatively small noise level $(\sigma = 0.05)$. Our goal is to compare \Mtrue with \Mfit and find a way to summarize how good the fitting result is.

\subsection{Element Baselines}
\label{sec 2.2}

To quantify how element choice influences recovery performance, we define three baseline element sets and evaluate results obtained by adding to or removing from these baselines. The low coverage baseline contains four elements that are most commonly studied and easily measured even in low-resolution spectra: C, Mg, Ca, and Fe. The medium (Mid) coverage baseline includes nine elements that are relatively accessible observationally with high resolution spectroscopy: C, Na, Mg, Al, Si, Ca, Ti, Fe, Ni. The high coverage baseline is all the elements that have been measured in literature in any metal-poor star, referring from \texttt{JINAbase}\footnote{\url{https://jinabase.pythonanywhere.com/}}, and not as upper limit: C, N, O, Na, Mg, Al, Si, K, Ca, Ti, V, Mn, Fe, Co, Ni \citep{Abohalima_2018}. Within our framework, this represents the theoretical best performance.
Following \cite{Heger_2010} and recommendations from their \texttt{starfit}\footnote{\url{https://starfit.org}}  website, we exclude Sc, Cr, Cu, and Zn from the baselines.
We still include these four elements in the single-element add/remove tests described in Section \ref{sec 2.4} and report their inferred importance in Table \ref{tab:1} for completeness. However, we recommend interpreting their importance with additional caution relative to other elements, given known theoretical and modeling limitations for these species in HW10 (e.g., Sc and Cu are treated as lower bounds; \citealt{Heger_2010}), making the ranking results less trustworthy.

\begin{figure}[h]
    \centering
    \includegraphics[width=\linewidth]{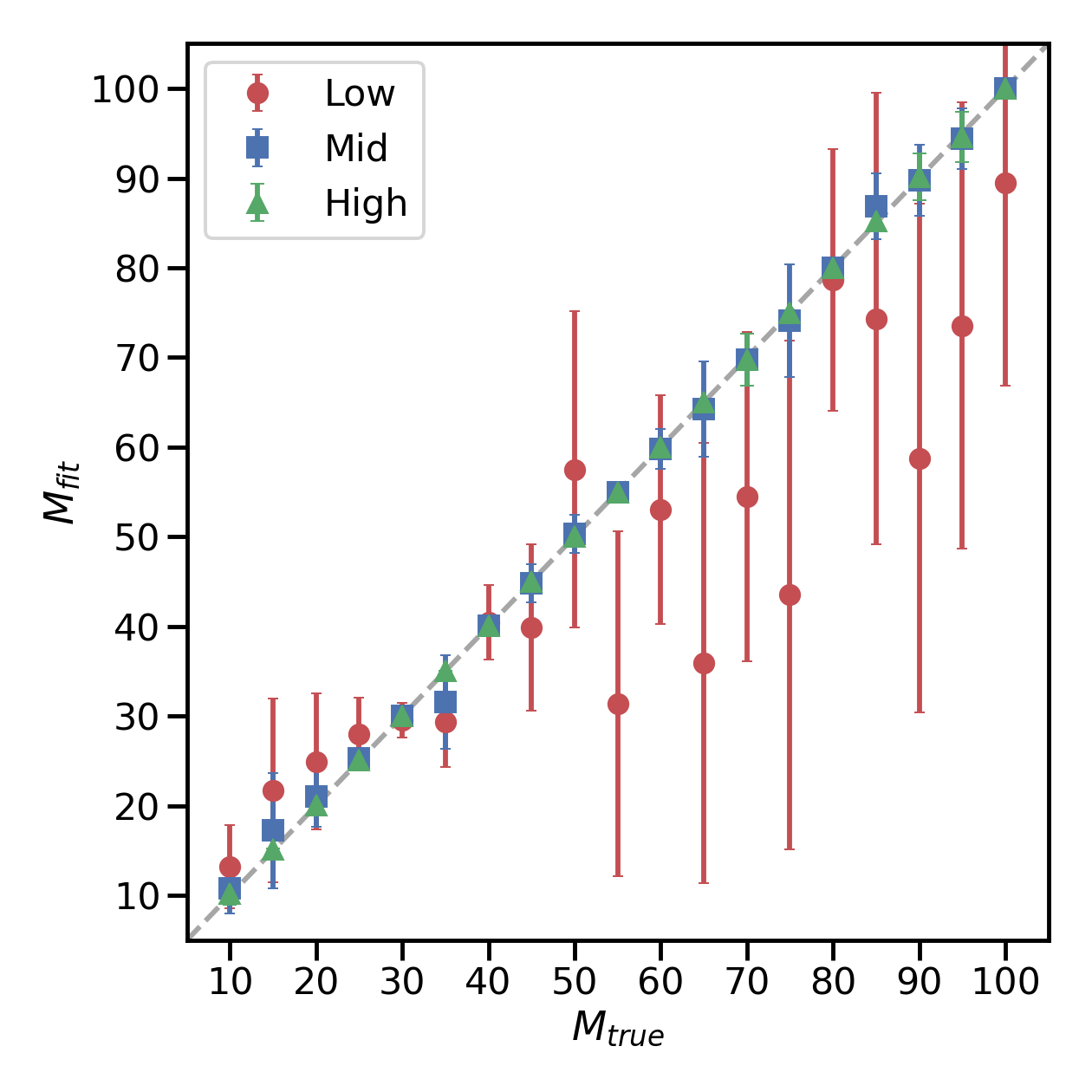}
    \caption{Recovery of Pop III star mass from mock observations under three baseline element sets: Low coverage (C, Mg, Ca, Fe), mid coverage (C, Na, Mg, Al, Si, Ca, Ti, Fe, Ni), and high coverage (C, N, O, Na, Mg, Al, Si, K, Ca, Ti, V, Mn, Fe, Co, Ni). Each point shows the mean and standard deviation of the 100 recovered masses \Mfit obtained from independent mock observations of a single HW10 model with the corresponding \Mtrue, \Etrue$=0.3$, and \Xtrue$=0$, and noise level $\sigma=0.2$; the gray dashed line denotes the one-to-one relation. The high coverage baseline places nearly all points on the dashed line, indicating near-perfect mass recovery in this process, whereas the low coverage baseline shows substantially larger scatter.
}
    \label{fig:1}
\end{figure}

Figure \ref{fig:1} presents fits for the three baselines at $E = 0.3 \text{B}$ and zero mixing with noise level $\sigma=0.2$ dex. We adopt $\sigma=0.2$ dex as a representative uncertainty for abundance measurements in metal-poor stars, and use it as our fixed noise level throughout this work. For simplicity and to allow controlled comparisons across element sets, we apply a constant $\sigma$ to all elements, treating it as an effective observational uncertainty.
Each point shows the mean and standard deviation of the 100 realizations of \Mfit associated with the corresponding \Mtrue. Under perfect recovery, the points would lie on the gray dashed one-to-one line. The low coverage baseline shows both substantial deviations from the dashed line and large dispersions, indicating inaccurate and unstable mass recovery at this noise level. We tested reducing the fixed observational noise to $\sigma=0.1$ and $\sigma=0.05$ and it improves overall recovery quality, but the result remains qualitatively insufficient under the same metric-based evaluation used throughout this work. The medium coverage baseline performs markedly better, with most points remaining close to the one-to-one relation, though residual offsets and uncertainties persist, especially at higher masses. The high coverage baseline yields results that are nearly perfect within a 1\% error, placing almost all points on the dashed line with relatively small scatter. These results generalize to other energy and mixing, which indicates that Population III yield fitting with all measurable elements can provide a reliable approach to mass recovery and IMF measurement.

\subsection{Quality}
\label{Sec 2.3}
\begin{figure}[h]
    \centering
    \includegraphics[width=\linewidth]{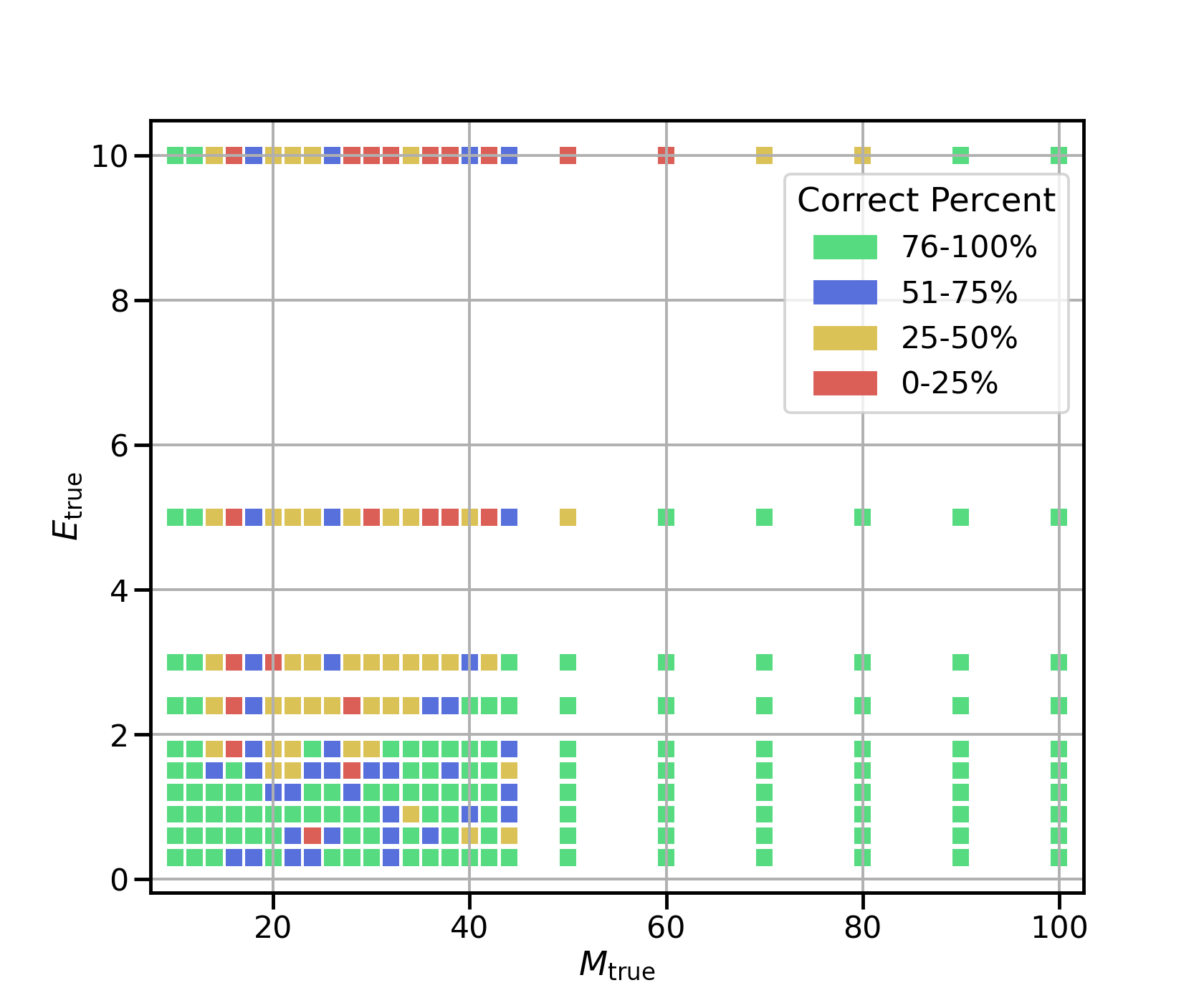}
    \caption{Demonstration of the metric calculation on the $M_{true}$--$E_{true}$ grid at a fixed mixing $X$. For each HW10 model (\Mtrue, \Etrue), 100 mock observations are generated and fit to recover 100 \Mfit. Points are colored by the fraction of correct recoveries, where a recovery is correct if $\delta = \left | (M_{true} - M_{fit}) / M_{true} \right | < 0.1$: green $76-100\%$, blue $51-75\%$, yellow $26-50\%$, red $0-25\%$. The frame-level metric is $d = (g + 0.5b - 0.5y - r)/N$, where $g, b, y, r$ are counts of points in each color and N is the number of points; $d$ ranges from -1 to 1, with higher values indicating better recovery. Example shown: \Xtrue = 0 and the medium coverage baseline element set. Only part of the grid is displayed for clarity.
}
    \label{fig:2}
\end{figure}

To enable quantitative comparison across element inputs, we construct a numerical metric that summarizes fit quality. For each element set, we compress the mass recovery result at fixed mixing into a single numeric score $d$, and then average over the 14 mixing frames to obtain \dbar for comparison across element sets. Figure \ref{fig:2} demonstrates the procedure. For a fixed mixing and a chosen element set, we go through the mass-energy grid of HW10 models. Each point corresponds to 100 mock observations generated from the original model with \Mtrue and \Etrue. For clarity, only part of the grid is shown in figure \ref{fig:2}. Across the full grid, there are 14 discrete mixings, so each fixed-mixing frame contains  16800/14 = 1200 points.

At each grid point, we evaluate the 100 \Mfit values from the 100 mock realizations with $\sigma$ = 0.2 and count how many are correct, where a recovery is correct if $\delta = \left | (M_{true} - M_{fit}) / M_{true} \right | < 0.1$. We encode the percentage correct by color: green $(76-100\%)$, blue $(51-75\%)$, yellow $(26-50\%)$, red $(0-25\%)$. To summarize a fixed-mixing frame, we compute $d = (g + 0.5b - 0.5y - r) / N$, where $g, b, y, r$ are the numbers of points in the green/blue/yellow/red categories and $N$ is the total number of points in the frame ($N = 1200$ in a full frame). This normalizes $d$ to $[-1, 1]$. More green/blue points increase $d$ and indicates a better recovery, while more yellow/red points decrease $d$ and means a worse recovery.

To summarize performance across mixing, we average $d$ over the 14 frames to obtain \dbar, which we use as the primary summary statistic for comparing element sets.

We also test alternative metrics and correctness definitions, such as reweighting the category counts in the $d$ or adopting an absolute mass error cut $\left | (M_{true} - M_{fit}) \right | < 2 M_\odot $, and observe no significant change in element rankings or relative importance. The metric we define is thus somewhat arbitrary, but sufficient for our purpose. 


\subsection{Element Importance}
\label{sec 2.4}
We assess element importance by adding or removing one element from a baseline and repeating the entire process: generate 100 mock observations across the HW10 grid and calculate \dbar. The noise level is fixed to $\sigma$ = 0.2.

\begin{figure*}
    \centering
    \includegraphics[width=\linewidth]{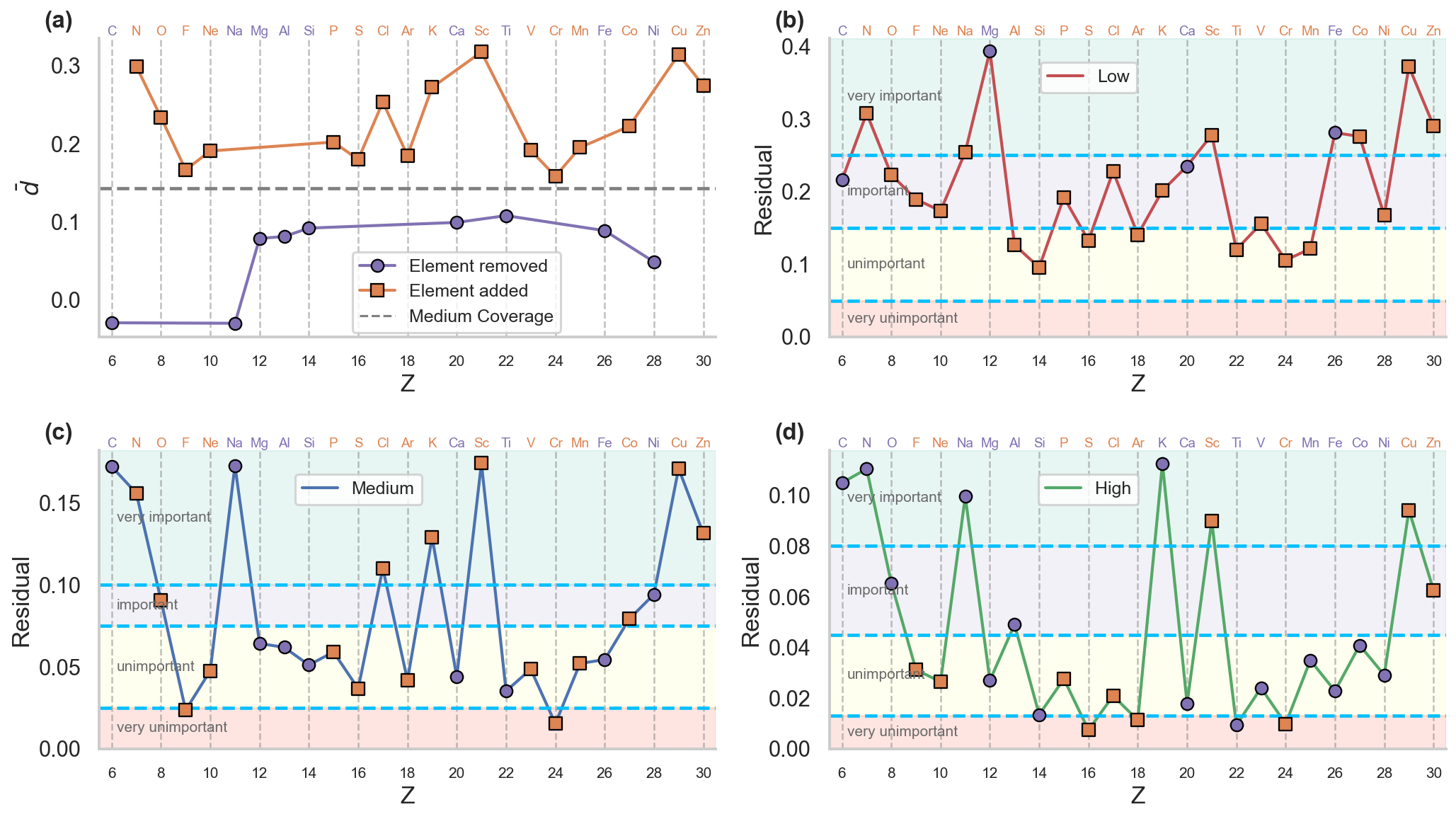}
    \caption{Element-by-element perturbations of the baselines. (a) Medium baseline: \dbar versus atomic number $Z$ when a single element is modified. The dashed line is the unmodified medium-baseline \dbar. Purple circle markers indicate the element is removed from the baseline; orange square markers indicate an element is added to the baseline. Greater differences from the dashed line imply a larger effect on the metric. (b-d) Low (red), medium (blue), and high (green) coverage baselines: $\Delta \bar{d} = |\bar{d}_{modified} - \bar{d}_{baseline}|$ versus $Z$. Horizontal dashed lines show the thresholds used to group elements into importance tiers. All results use 100 mocks per model and noise level $\sigma$ = 0.2; \dbar is averaged over the 14 mixing values.
}
    \label{fig:3}
\end{figure*}

Figure \ref{fig:3} (a) shows \dbar versus atomic number $Z$ when a single element is added to or removed from the medium baseline. The dashed line marks the medium-baseline \dbar. A circle purple point means the element is removed; a square orange point means the element is added. Points further from the dashed line indicate larger changes in the metric and thus greater importance for mass recovery.

To quantify this, we compute the absolute change $\Delta \bar{d}$ between each point and the dashed line, where the dashed line denotes the $\bar{d}$ value of the baseline used in that panel. Figures \ref{fig:3}(b-d) present $\Delta \bar{d}$ for the low, medium, and high baselines. A larger $\Delta \bar{d}$ indicates a more important element. Using horizontal dashed thresholds, we categorize elements into four groups: very important, important, unimportant, and very unimportant. These thresholds are chosen by visual inspection to separate the apparent clusters in $\Delta \bar{d}$ within each baseline panel and therefore differ among subplots. For elements near a boundary, we place the horizontal cuts by also considering the $\Delta d$ values in each mixing frame: an element is assigned to the higher tier if its $\Delta d$ is relatively large across the individual mixing frames. Table \ref{tab:1} lists the resulting categories, and the identities of the top-tier elements are stable to reasonable changes in the metric and correctness definition.

\begin{table*}[t]
\centering
\small
\setlength{\tabcolsep}{4pt}
\renewcommand{\arraystretch}{1.7}
\begin{tabular}{|l|l|l|l|}
\hline
 & Low & Medium & High \\ \hline
Very Important &
N, Na, Mg, \textcolor{red}{Sc}, Fe, Co, \textcolor{red}{Cu}, \textcolor{red}{Zn} &
C, N, Na, Cl, K, \textcolor{red}{Sc}, \textcolor{red}{Cu}, \textcolor{red}{Zn} &
C, N, Na, K, \textcolor{red}{Sc}, \textcolor{red}{Cu} \\ \hline
Important &
\parbox[c]{3.5cm}{C, O, F, Ne, P, Cl, K, \\[1.5pt] Ca, V, Ni} &
\parbox[c]{2.5cm}{O, Co, Ni} &
\parbox[c]{2.5cm}{O, Al, \textcolor{red}{Zn}} \\ \hline
Unimportant &
\parbox[c]{3.5cm}{Al, Si, S, Ar, Ti, \textcolor{red}{Cr}, Mn} &
\parbox[c]{4.5cm}{Ne, Mg, Al, Si, P, S, Ar, \\[1.5pt] Ca, Ti, V, Mn, Fe} &
\parbox[c]{4.5cm}{F, Ne, Mg, Si, P, Cl, Ca, \\[1.5pt] V, Mn, Fe, Co, Ni} \\ \hline
Very unimportant &
\parbox[c]{3.5cm}{} &
\parbox[c]{2.5cm}{F, \textcolor{red}{Cr}} &
\parbox[c]{2.5cm}{S, Ar, Ti, \textcolor{red}{Cr}} \\ \hline
\end{tabular}
\caption{Element importance categories derived from Figure \ref{fig:3} (b-d). Elements are grouped as Very Important, Important, Unimportant, or Very Unimportant based on $\Delta \bar{d}$, with thresholds given by the horizontal dashed lines in those panels. Elements shown in red (Sc, Cr, Cu, Zn) are excluded from our baseline sets but are included in the single-element add/remove tests and reported here for completeness. Their inferred rankings should be interpreted with additional caution, given known theoretical and modeling limitations for these species in HW10, making these entries less trustworthy than other elements.}
\label{tab:1}
\end{table*}

\subsection{Mass Uncertainty}
\begin{figure}[h]
    \centering
    \includegraphics[width=\linewidth]{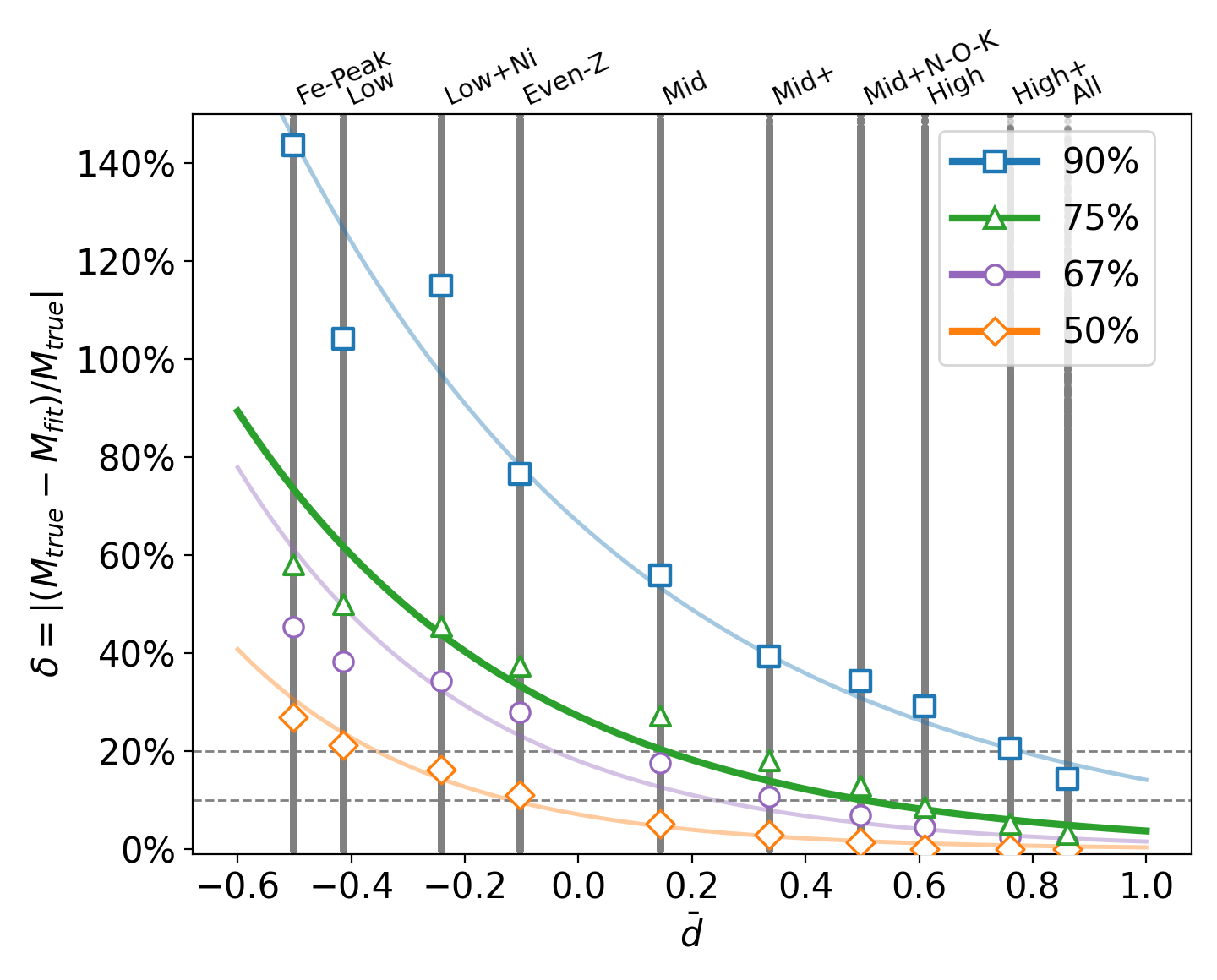}
    \caption{Mapping between the fit quality metric $\bar{d}$ and the fractional mass error $\delta = \lvert(M_{true}-M_{fit})/M_{true}\rvert$. At the $\bar{d}$ attained by ten representative element sets (Fe-Peak, Low, Low+Ni, Even-Z, Mid, Mid+, Mid+N-O-K, High, High+, All), scatter points show all $\delta$ values across the HW10 grid (16{,}800 models $\times$ 100 mocks $= 1.68\times10^{6}$ points per set). Symbols mark the 50th, 68th, 75th, and 90th percentile thresholds at each $\bar{d}$, and the smooth curves are exponential fits to these thresholds. The 75th-percentile curve is highlighted (thicker line) since it is used to compute $\delta_{75}$ in Figure~\ref{fig:5}. }
    \label{fig:4}
\end{figure}

Our goal in this subsection is to calibrate the summary metric \dbar to the expected uncertainty in progenitor mass recovery. This provides a quantitative mapping from fit quality to IMF measurement uncertainty, enabling observational planning: given a target percentage mass error and a desired success rate, one can read off the required \dbar from Figure \ref{fig:4} and then, in Section \ref{Results}, identify which element sets typically achieve that $\bar{d}$ with Figure 5. Here we define the percentage mass error as $\delta = |(M_{\rm true}-M_{\rm fit})/M_{\rm true}|$ and interpret the percentile ($p\%$) curves in Figure \ref{fig:4} as follows: at a given $\bar{d}$, the $p\%$ curve gives the value $\delta_p$ such that $p\%$ of mock observations satisfy percentage mass error $\delta \le \delta_p$. We highlight $\delta = 10\%$ and $20\%$ as practical accuracy targets, and use the medium coverage baseline as a reference point: its $75\%$ curve corresponds to $\delta \approx 20\%$, which we take as a minimal “trustworthy” regime for downstream IMF inference; accordingly, element sets with \dbar exceeding the medium coverage baseline can yield reliably accurate mass recovery.

Figure \ref{fig:4} relates \dbar to the percentage error in mass recovery, $\delta$. For each selected \dbar drawn from the baselines and several representative element combinations, we pool the distribution of $\delta$ across the HW10 grid under the same setup as in Sections 2.1-2.4 ($\sigma$ = 0.2 noise; all masses and energies; all mixings aggregated). Concretely, each of the 16,800 HW10 models generates 100 mock observations, yielding 1,680,000 \Mfit values, and thus 1,680,000 $\delta$ values, for every \dbar shown on the plot. We then summarize the distribution at that \dbar by its 50th, 67th, 75th, and 90th percentiles and fit simple exponential curves to these percentile threshold markers. The colored curves in Figure \ref{fig:4} therefore provide a smooth, monotonic mapping.

To anchor this mapping to observing choices, we evaluate ten representative element sets: the low, medium, and high coverage baselines; low coverage baseline with Ni (low+Ni); an iron-peak set (Ti, V, Mn, Fe, Co, Ni); an even-Z set (C, Mg, Si, Ca, Ti, Fe, Ni); medium coverage baseline augmented with Sc and Cr (Mid+); medium coverage baseline with N, O, and K (Mid+N-O-K); high coverage baseline augmented with Sc, Cr, and Zn (High+); and all 25 elements. For each set we plot, at its achieved \dbar, all individual $\delta$ values as semi-transparent scatter points. The distinct marker shapes along each cloud mark the 50th, 68th, 75th, and 90th percentile thresholds, and the smooth curves are exponential fits to these percentile markers. Horizontal dashed guides at $\delta = 10\%$ and $20\%$ mass precision highlight commonly used error targets.

Figure \ref{fig:4} can be interpreted in two ways. For a fixed element set, the percentile curves can be read vertically: the $p$-th percentile curve gives the value $\delta_p$ such that $p\%$ of all mock observations across the HW10 grid recover masses with $\delta \le \delta_p$. For example, for the Mid set, the plot indicates that roughly $75\%$ of mock observations achieve $\delta \lesssim 20\%$, while $90\%$ achieve $\delta \lesssim 60\%$. Alternatively, the mapping can be read as an analysis target: for a desired accuracy threshold (e.g., $\delta = 10\%$ or $20\%$), one reads horizontally to the intersection with a chosen percentile curve and then projects down to the corresponding $\bar{d}$ requirement, which can be compared against the $\bar{d}$ values attained by candidate element sets. For example, achieving a $50\%$ chance that $\delta \le 10\%$ requires only the Even-Z set, with $\bar{d} \approx -0.035$; requiring the same $\delta \le 10\%$ target with $67\%$ success calls for Mid+ with $\bar{d} \approx 0.339$; and pushing to a $75\%$ success rate at $10\%$ accuracy requires the high baseline, with $\bar{d} \approx 0.623$.

\section{Results} \label{Results}

We evaluate \dbar for multiple element combinations beyond single-element add/remove tests, aiming to compare reasonable choices with the low/mid/high coverage baselines. The setup is consistent with Section \ref{Methods}: 100 mocks per HW10 model, noise level $\sigma=0.2$, and \dbar averaged over the 14 mixing prescriptions.

\begin{figure*}[htbp]
    \centering
    \includegraphics[width=\linewidth]{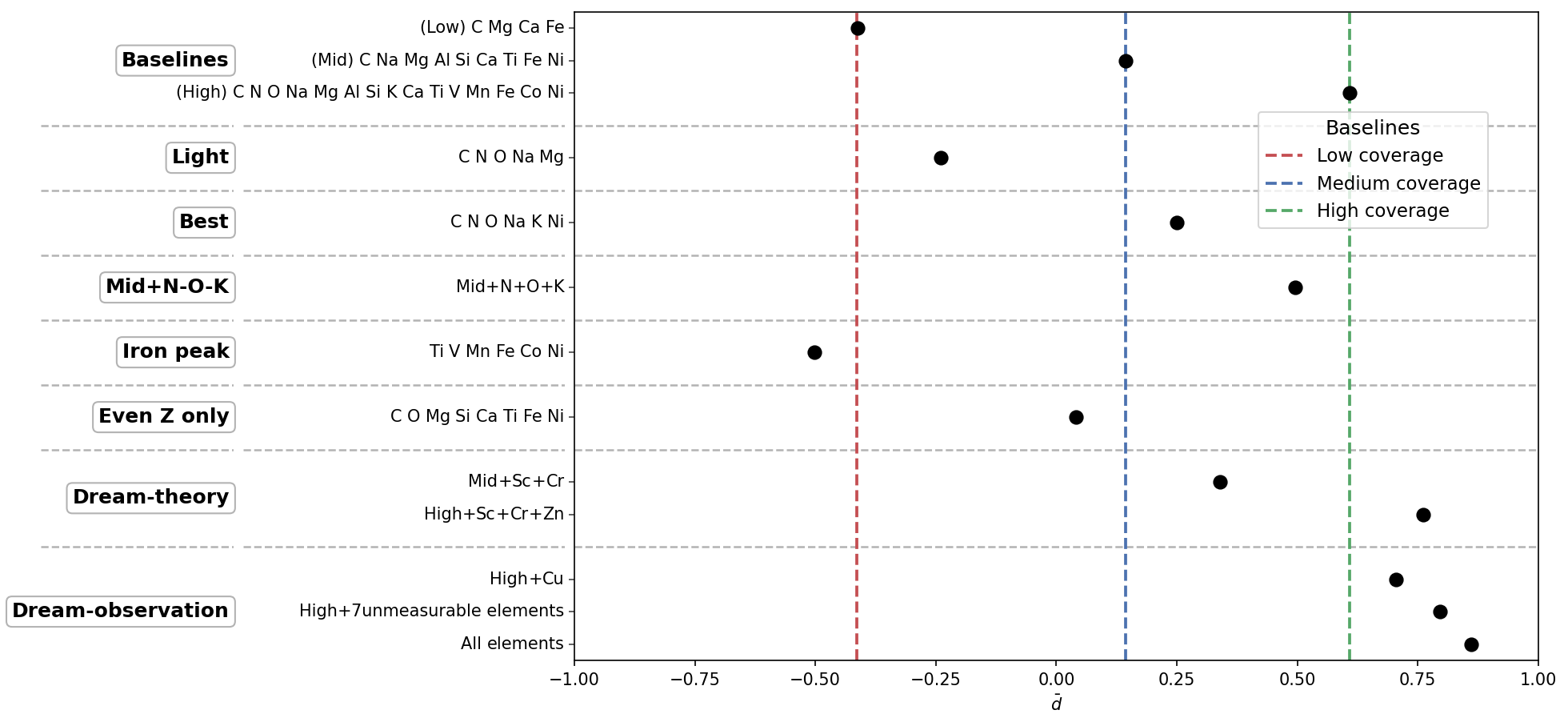}
    \caption{Summary of mass recovery performance across different element sets. Each point corresponds to a tested element combination grouped by the category labels listed on the left. For each element set, we report its \dbar value together with the corresponding $\delta_{75}$ value, where $\delta_{75}$ denotes the 75th-percentile fractional mass error. The dashed reference lines indicate the low (red), medium (blue), and high (green) coverage baselines. In this framework, element sets lying to the right of the medium coverage baseline are considered as good mass recovery performance.}
    \label{fig:5}
\end{figure*}

Figure \ref{fig:5} shows one \dbar point for each combination. The three vertical dashed lines mark the low, medium, and high coverage baseline values; combinations to the right of a given line outperform that baseline, while those to the left underperform it. Labels on the y-axis group related combinations for ease of comparison. We additionally explored a broader set of element combinations; the complete version of Figure 5 is provided in the Appendix.

Baseline behavior is as expected: low performs poorly, medium is moderate, and high is strong. If every low-metallicity star were observed at high coverage, the IMF could be constrained to high precision when paired with the $\bar{d} \mapsto \delta$ mapping in Figure \ref{fig:4}. The result for $\delta_{75\%}$ is at the top of Figure \ref{fig:5}.
Adding one important element to the low coverage baseline (Low+) yields only a small increase in \dbar, which is not enough for a precise IMF measurement. By contrast, adding N, O, and K to the Mid set (Mid+N-O-K) produces a large positive shift in \dbar, consistent with our single-element analysis, shown in Figure \ref{fig:3}, identifying N, O, and K as important for mass recovery. 

The light-only choice performs poorly, showing that light elements alone are insufficient, indicating that iron peak elements, which carry explosion information, are necessary. Similarly, the iron peak only sets are also poor, implying that both light and Fe-peak elements are required. The even-Z only choice performs poorly as well, reinforcing the importance of odd-Z elements such as N, Na, and K.

Finally, the ``dream'' sets show the possibilities if additional species can be modeled and measured robustly. These are implemented as medium coverage baseline augmented with Sc and Cr, and high coverage baseline augmented with Sc, Cr, and Zn. They are the same as Mid+ and High+ sets referenced in Figure \ref{fig:4}. Both shifts move their points distinctly to the right of the medium and high coverage baselines. On the observational side, adding currently non-observable elements (F, Ne, P, S, Cl, Ar, and Cu) also improves the fitting result (``dream-observation''). Among them, adding Cu to the high coverage baseline produces the largest rightward shift in \dbar, but Cu also carries theoretical and modeling issues, as explained in Section \ref{sec 2.2}. The last row of Figure \ref{fig:5}, which includes all 25 elements from C to Zn, is the practical upper limit in this setup.

\section{Discussion and Conclusion} \label{Conclusion}
We present a controlled, quantitative assessment of progenitor mass recovery from Pop III yield fitting. Starting from the HW10 grid, we generated mock observations, fit them across the full HW10 grid, summarizing each fit into a quantitative metric \dbar. We map \dbar to percentage mass recovery error $\delta$ via exponential curves, providing a direct translation from fit quality to IMF precision. Applied to systematic add/remove-one-element tests and to physically motivated element sets, this framework yields guidance for observation planning: choose a target $\delta$ and a success percentile $p$ and compare with element combinations in Figure \ref{fig:5} to determine whether it meets or exceeds the threshold.

Several robust conclusions emerge. The most consistently very important elements are C, N, Na, and K, with O, Al, Co, and Ni also important. A practical takeaway is that the current measurable elements set, corresponding to our high coverage baseline, is already sufficient for the IMF calculation at useful precision. Notably, K emerges as a very important contributor to mass recovery even though it is not often measured in many metal-poor star abundance analysis, and our tests indicate that including K can substantially improve mass recovery performance. More broadly, mass recovery performance depends on complementarity: both light and iron peak elements are required, and odd-Z elements are critical in mass recovery.

Our element importance ranking aligns with most of the prior Pop III fitting studies. We find C, N, O, and Na are the most consistently important elements for mass recovery, which agree with \citet{Tominaga_2014, Bessell_2015, Placco_2015} and \citet{Ishigaki_2018}. Since these elements are individually important, ratios constructed from them, such as those involving C/N/O and Na/Mg/Al, are informative, which is consistent with \cite{Tominaga_2014} and \cite{Ishigaki_2018}.
Our results also partially agree with \citet{Hansen_2011, Ishigaki_2014, Ishigaki_2018} and \citet{Jiang2025} that C and Na emerge as consistently informative across element sets. For other elements emphasized in these studies, we find a clear baseline dependence: the influence of Mg, Fe, and Co is concentrated in the low coverage regime, where they provide substantial leverage when only a small set of abundances is available, but they are unimportant for the medium and high baselines, presumably because the information is covered by other elements. By contrast, Al, also highlighted in several of these works, shows the opposite behavior in our add/remove tests, with its contribution becoming most pronounced for the high coverage baseline.
A key contrast is that several works highlight Si, V, and Mn as potentially informative \citep{Ishigaki_2014,Ishigaki_2018,Hansen_2011,Jiang2025}, whereas in our single-element add/remove tests these elements are consistently ranked as unimportant.
We also find Sc, Cu, and Zn to be important in our add/remove tests, which agree with \citet{Hansen_2011, Ishigaki_2014, Ishigaki_2018}, but we regard their rankings as less trustworthy given known modeling limitations for these species in HW10 models.
Finally, our results highlight K as a consistently very important contributor to mass recovery, a point that is not emphasized in the Pop III fitting studies discussed above.

Elements that are ranked as unimportant in low coverage baseline can still yield substantial absolute improvements in mass recovery when added. Therefore, in the low-coverage regime, we recommend incorporating any additional element abundances that are available with reasonable precision, even if they are ranked as unimportant in low coverage baseline. In general, adding more elements tends to improve performance, but this is not guaranteed for every situation. For some specific mixing in medium and high coverage baseline, adding one element to the baseline decreases the metric, or removing one element increases the metric. However, these only yield small deviations in \dbar with the order of magnitude of $10^{-3}$ dex, which correspond to negligible changes in $\delta$ under our mapping.

We also tested the robustness of our conclusions to how the metric is constructed. Alternative definitions of correctness and different color groupings leave the overall importance hierarchy intact, with only small shuffling between the ``very important'' and ``important'' tiers. Thus, the identity of the high-leverage elements is stable with respect to the specific metric that we chose.

This study has three key limitations. First, we report a single global element ranking averaged over the full mass-energy grid. It could be beneficial to repeat the importance analysis within smaller subregions of parameter space to test the stability of our results in specific mass/energy ranges of interest. Second, our analysis is restricted to progenitor mass recovery; extending the same end-to-end mock-and-refit framework to explosion energy will clarify which elements primarily constrain energy versus mass. Third, all results are based on a single core-collapse supernova yield grid. Repeating the full study across multiple yield models will assess how robust the inferred importance conclusion remains under different modeling assumptions. These extensions will sharpen observational recommendations and strengthen the robustness of IMF inference. Nevertheless, under the assumptions explored here, the set of elements that are measurable today is already sufficient to deliver informative constraints on the Pop III IMF.

\newpage

\begin{acknowledgments}

Z.Z. and A.P.J. acknowledge support from the National Science Foundation under grants AST-2206264 and AST-2510795. A.P.J. acknowledges support from the Alfred P. Sloan Foundation.
The work of V.M.P. is supported by NOIRLab, which is managed by the Association of Universities for Research in Astronomy (AURA) under a cooperative agreement with the U.S. National Science Foundation.

\end{acknowledgments}

\begin{figure*}[htbp]
    \centering
    \includegraphics[width=\linewidth]{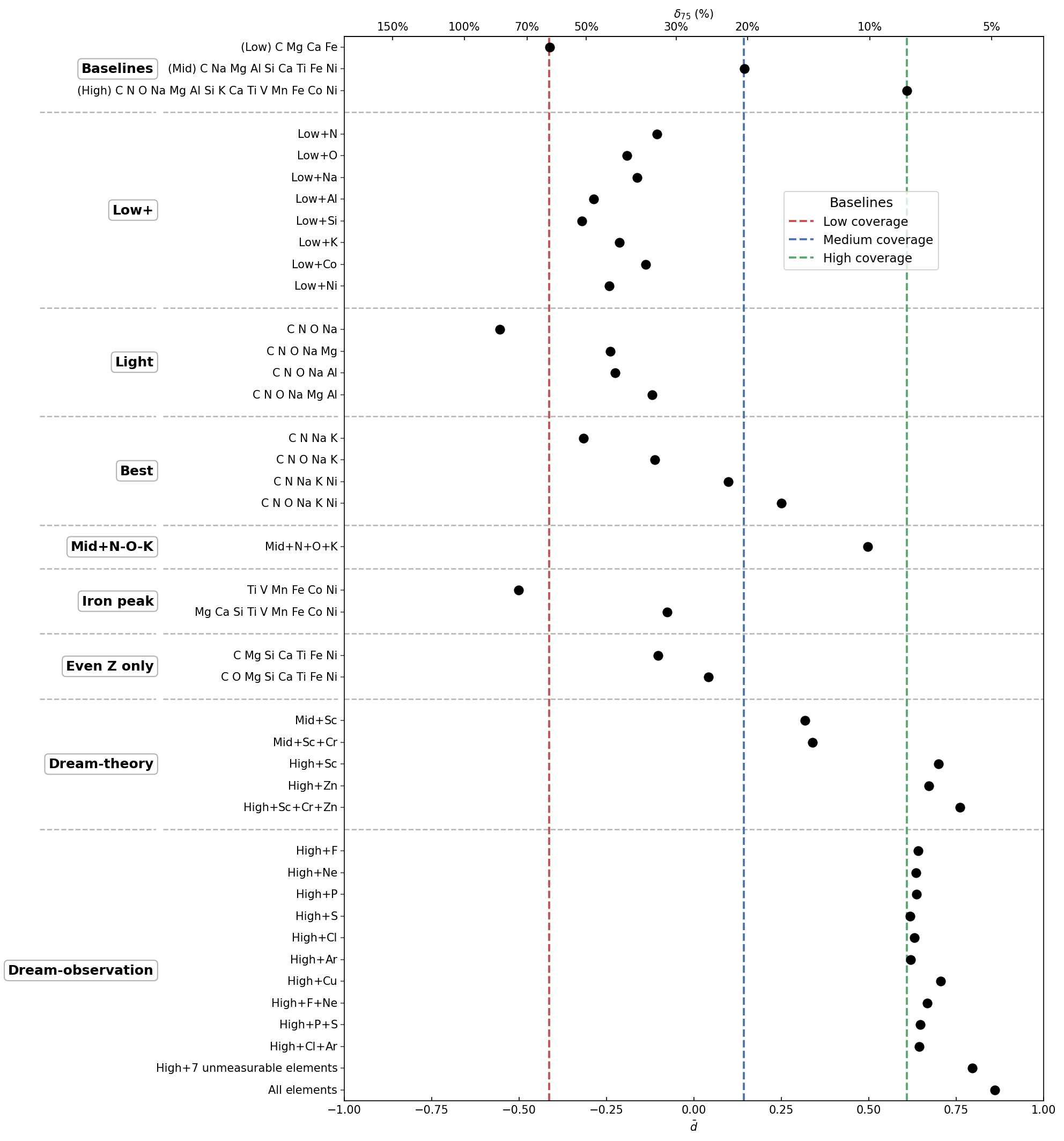}
    \caption{Full version of Figure 5 including all tested element combinations, shown with the low, medium, and high coverage baseline reference lines.}
    \label{fig:6}
\end{figure*}

\appendix
Figure \ref{fig:6} presents the full version of Figure \ref{fig:5}, including all additional element combinations tested beyond those highlighted in the main text. The same baseline reference lines are shown to facilitate direct comparison across categories.

\newpage


\end{document}